# Constitution and Model:
# Bohr's Quantum Theory and Imagining the Atom

Giora Hon* and Bernard R. Goldstein**

> There seem good reasons for believing that radio-activity is due to changes going on within the atoms of the radio-active substances. If this is so then we must face the problem of the constitution of the atom, and see if we can imagine a model which has in it the potentiality of explaining the remarkable properties shown by radio-active substances.
>
> J. J. Thomson (1904)[1]

> It could be that I've perhaps found out a little bit about the structure of atoms. ... If I'm right, it would not be an indication of the nature of a possibility [marginal note in the original: "i.e., impossibility"] (like J. J. Thomson's theory) but perhaps a little piece of reality.
>
> N. Bohr (1912)[2]

**Abstract**: Bohr's theory has roots in the theories of Ernest Rutherford and Joseph J. Thomson on the one hand, and that of John W. Nicholson on the other. We note that Bohr neither presented the theories of Rutherford and Thomson faithfully, nor did he refer to the theory of Nicholson in its own terms. Bohr's contrasting attitudes towards these antecedent theories is telling and reveals his philosophical disposition. We argue that Bohr intentionally avoided the concept of model as inappropriate for describing his proposed theory. Bohr had no problem in referring to the works of others as "models", thus separating his theory from previous theories. He was interested in uncovering "a little piece of reality".

**Kew words**: Rutherford; J. J. Thomson; Nicholson; atom-model; Planck's quantum of action.

---

* Department of Philosophy, University of Haifa, Israel. E-mail: hon@research.haifa.ac.il.
** Dietrich School of Arts and Sciences, University of Pittsburgh, USA. E-mail: brg@pitt.edu.

[1] Thomson (1904b), p. 92.

[2] A letter from Niels Bohr to his brother, Harald, dated 19 June 1912. Quoted in Heilbron and Kuhn (1969), p. 238.

In examining a large number of texts by prominent physicists from the late nineteenth and the first decade of the twentieth centuries we have found that the term "model" at that time referred to a representation based on mechanical principles of a physical system. Against this background we claim that, contrary to the widely accepted view, Niels Bohr did not intend to develop a model of the atom in his Trilogy. Rather, he was interested in the real thing, a theory that accounts for experimental results by means of real entities; to be specific, his goal was to lay bare the constitution of the atom as the title of the Trilogy indicates. Our paper is then a contribution towards a history of modeling in twentieth century physics.[3]

Bohr's theory has roots in the theories of Ernest Rutherford and Joseph J. Thomson on the one hand, and that of John W. Nicholson on the other. We note that Bohr neither presented the theories of Rutherford and Thomson faithfully, nor did he refer to the theory of Nicholson in its own terms. Bohr's contrasting attitudes towards these antecedent theories is telling and reveals his philosophical disposition.

The first reaction to Bohr's Trilogy from outside the Rutherford circle came from Arnold Sommerfeld. Early in September 1913 Sommerfeld wrote to Bohr: "I thank you very much for sending me your highly interesting paper, which I had already read in the Phil. Mag."[4] Sommerfeld then expressed some skepticism concerning the application of atomic models [*Atommodellen*], and wondered whether Bohr would apply his atom model to the Zeeman effect. Sommerfeld, a mathematical physicist, understood Bohr's theory in terms of modeling.[5] By 1914 Bohr's theory was accepted as a model in the literature. However, a careful reading of the Trilogy leads us to claim that this was not the way Bohr presented his theory. In fact, we are persuaded that the concept of model was extended as a result of including Bohr's theory in the category of model, but this development was not due to Bohr.

In the Trilogy Bohr wished to distinguish his theory from the conceptions of Rutherford and Thomson. Right at the outset Bohr explicitly called the theories of Rutherford and Thomson "atom-model" while considering his own theory an attempt at uncovering the constitution of the atom. In so doing, Bohr did not accurately report the works of his two mentors. Since his theory includes Planck's hypothesis and theirs did not, something fundamental separated the old theories from his own. To be sure, there was a precedent, namely Nicholson had proposed a theory in 1912 that included Planck's hypothesis in which Nicholson invoked the expression "model atom". Moreover, Nicholson called his theory "model" while seeking, as he put it, "the constitution of the solar corona".[6] But Bohr never refers to Nicholson's theory as "model"; in fact, he systematically calls Nicholson's proposal "theory". We find that Bohr is consistent in his claim for "constitution"—he does not propose a model; the Trilogy was not intended to describe a representation of the atom. Bohr, we argue, took the concept of orbit from Thomson and the nucleus from Rutherford. He also noticed that Nicholson included Planck's quantum of action together with the constraint of the constancy of angular momentum—all of which Bohr considered "real". Sommerfeld's immediate response indicates, however, that despite Bohr's apparent intention to assert that his theory dealt with constitution and not with modeling, the theory was quickly perceived as a model.

---

[3] Cf. Hon and Goldstein (2012).

[4] Sommerfeld to Bohr, 4 September 1913, in *BCW* 2 (1981), pp. 123, 603.

[5] For Sommerfeld's views in 1911 on models, see Sommerfeld (1912), p. 124: "As for me, I prefer a general hypothesis for *h* rather than specific atomic models." ("Quant à moi, je préfère une hypothèse générale sur *h* à des modèles particuliers d'atomes.") Cf. Sommerfeld (1911), p. 1066 (quoted in n. 21, below).

Words count—they are, after all, markers of concepts. This statement should not be dismissed lightly as an inconsequential truism. When analyzing a concept in the history of science there is the tendency to make the comment (implicitly), "What's in a word?" We take a different approach, namely in our view attention should be focused on the usages of terms and the changes in their meanings. We argue that linguistic usages reflect philosophical dispositions and in this paper we explore these underlying dispositions. The issue we address has to do with conceptual frameworks, taking words seriously as markers of concepts. In 1913 "modeling" had a specific meaning, namely a mechanical or electrical system for representing another physical system, e.g., the ether. We argue that Bohr intentionally avoided the concept of model as inappropriate for describing his proposed theory.

Bohr began his pathbreaking paper of 1913 with a reference to the surprising experimental result of large angle scattering of α rays by matter, obtained at Rutherford's laboratory in Manchester. Rutherford explained the results of this experiment by proposing an atomic structure in his paper of 1911, "The structure of the atom."[7] Rutherford thus echoed Thomson (1904), "On the structure of the atom."[8] Bohr, by contrast, titled his paper of 1913, the famous Trilogy, "On the constitution of atoms and molecules."[9] Bohr speaks, then, of "constitution", while associating Rutherford with "a theory of the structure of atoms". He goes on to call Rutherford's theory an "atom-model" which was not Rutherford's terminology in 1911. In the next paragraph Bohr considers Thomson's proposal an "atom-model". As in the case of Rutherford, this is not Thomson's terminology in his paper of 1904.[10] Bohr then calls Thomson's atom-model a "theory" and, according to Bohr, it was designed to avoid instability in combining positive electrification with fast moving negatively charged particles, the electrons (called corpuscles by Thomson). Thus, "theory" and "model" are used for both Rutherford's and Thomson's conceptions of the atom, while these two authors refer to the

---

[6] Nicholson (1912), p. 677.

[7] Rutherford (1911).

[8] Thomson (1904a).

[9] Bohr (1913).

[10] However, it became common to call this theory a model. Cf., e.g., Rutherford (1906), pp. 2, 265, 267, and Nicholson (1912), p. 686.

atom as having some "structure".[11]

Bohr now turns to compare the atom-models of his mentors. The radius of Thomson's atom-model is in fact the radius of the positive sphere—the linear extension of the atom. However, such a length cannot be defined in terms of Rutherford's atom-model. The former model, it should be noted, was conceived mathematically for the purpose of studying the atom's stability,[12] while the latter was founded on experimental results.[13]

Against the contrast between these two atom-models, Bohr considers the physics of energy radiation. He summarizes the situation, remarking that

> whatever the alteration in the laws of motion of the electrons may be, it seems necessary to introduce in the laws in question a quantity foreign to the classical electrodynamics, *i.e.*, Planck's constant, or as it often is called the elementary quantum action.[14]

Interestingly, Bohr considers the quantum of action a quantity foreign to electrodynamics. Furthermore, it is noteworthy that in their respective theories of the atom neither Thomson nor Rutherford appealed to Planck's quantum of action. According to Bohr, Planck's quantum of action, together with the mass and charge of the particles, determines the size of the atom. And Bohr states that the purpose of his paper is to apply these ideas to Rutherford's atom-model which "affords a basis for a theory of the constitution of atoms".[15] This indicates that by introducing the quantum of action into Rutherford's atom-model, Bohr intended to discard modeling and move towards constitution.

What did Bohr mean by atom-model? A relevant precedent for this idea was well known to Bohr. In 1912 Nicholson, a British astrophysicist, proposed a theory of atomic structure which he then applied to stellar spectroscopy and the periodic table. In his paper, "The constitution of the solar corona," Nicholson remarks,

> The constant of nature in terms of which ... spectra can be expressed appears to be that of Planck in his recent quantum theory of energy. It is evident that the model atoms with which we deal have many of the essential characteristics of Planck's "resonators."[16]

---

[11] Bohr (1913), I, p. 1. Cf. *BCW* 2 (1981), pp. 529–531 (a letter from Bohr to Hevesy, dated 7 Feb. 1913).

[12] Thomson (1904a), pp. 255–256. The full title of this paper refers to stability.

[13] Rutherford (1911), pp. 670–671 and § 6, "Comparison of theory with experiments".

[14] Bohr (1913), I, p. 2.

[15] Bohr (1913), I, pp. 2–3.

[16] Nicholson (1912), p. 677.

So in 1912 Nicholson recognized the relevance of Planck's quantum of action, not only with respect to spectral phenomena associated with Planck's resonators, but also with respect to the model atom. Nicholson dealt mechanically with a model atom, as he called it, to which he applied Planck's quantum hypothesis. Moreover, he took for granted Rutherford's result, assuming an atomic structure that included a "positive nucleus". Nicholson further referred to Thomson's "atomic model" and the associated equations of stability and periods of oscillation.[17]

What was the goal of Nicholson's study when he introduced Planck's theory into the discussion? First and foremost, he wanted to test whether it is "in accord" with his own spectral theory.[18] If so, then

> the investigation will serve the double purpose of confirming the suggested origin of the spectra of astrophysics, and of giving Planck's theory an atomic foundation: a foundation of the kind which is now generally believed to be necessary, giving a concrete picture of the possible nature of a resonator.[19]

Nicholson sought to accommodate Planck's theory with other theories, notably his own. He did not consider it fundamental or that his theory had to be built upon that of Planck; no, Planck's theory was regarded as instrumental in confirming the feasibility of Nicholson's own theory. He used a mechanical rotator to set up the quantum condition for interpreting a few spectral lines observed in the Sun and nebulae. Moreover, he did not consider the emitted radiation in terms of quanta. Still, he remarked that according to Planck's theory, "interchanges of energy are not continuous, so that it is not possible to represent ultimate dynamics by sets of differential equations."[20] Nicholson was most likely aware of the possible fundamental nature of Planck's theory since he referred to Sommerfeld who a year earlier argued for the fundamental nature of Planck's quantum of action.[21]

---

[17] Nicholson (1912), pp. 683, 686.

[18] Nicholson (1912), p. 677.

[19] Nicholson (1912), pp. 677–678.

[20] Nicholson (1912), p. 677.

[21] Nicholson (1912), p. 679. Cf. Sommerfeld (1911), p. 1066: "I would rather prefer the reverse point of view: instead of explaining $h$ by recourse to the dimensions of molecules, one should regard the existence of molecules as a function and a consequence of the existence of an elementary quantum of action." ("Vielmehr möchte ich den umgekehrten Standpunkt bevorzugen: das $h$ nicht aus den Moleküldimensionen zu erklären, sondern die Existenz der Moleküle als eine Funktion und Folge der Existenz eines elementaren Wirkungsquantums anzusehen.")

While Nicholson had a significant influence on Bohr, it appears that Bohr sought to distance himself from the practice of modeling the atom. Nicholson's goal was to describe the constitution of the solar corona: modeling of the micro-level was intended to help in accounting for the macro-phenomena.[22]

In the Trilogy Bohr addresses Nicholson's theory in two places.[23] At first he records the excellent agreement between calculations based on this theory and the observed values. Bohr, however, raises a serious objection, namely that Nicholson considered systems in which the frequency is a function of energy, and such a system cannot emit a finite amount of homogeneous radiation. In fact, according to Bohr, such systems are unstable. Thus, there could be no coherent account of the Balmer and the Rydberg series. Later on in his paper Bohr withdrew some of the criticism. In this second phase of his response to Nicholson, Bohr appeals to the constraint that Nicholson had introduced, namely the universal constancy of angular momentum.[24] Bohr now recognizes that Nicholson had applied his theory in an extreme case (the solar corona) unlike Bohr's own analysis of the relatively simple spectral series of Balmer and Rydberg.[25] In brief, Bohr accepted Nicholson's introduction of Planck's quantum of action into atomic theory, but he did not address Nicholson's appeal to "model". While Bohr called Thomson's and Rutherford's theories of structure "models", he did not refer to Nicholson's theory as a model even though Nicholson himself called his theory "model". We ask, Why did Bohr avoid calling Nicholson's theory a model? To be sure, the difference between Nicholson's approach and Bohr's methodology is striking, for Bohr turned Planck's quantum of action into one of the two postulates of his theory.[26] In other words, Bohr reversed the way Nicholson introduced Planck's quantum of action. However, while the atomic theory of Thomson and that of Rutherford are essentially classical in that they do not include Planck's quantum of action, Nicholson's theory has components similar to Bohr's theory. In fact, Bohr reported to Rutherford early in 1913 that

> In his calculations, Nicholson deals, as I, with systems of the same constitution as your [Rutherford's] atom-model; and in determining the dimensions and the energy of the system he, as I, seeks a basis in the relation between the energy and the frequency suggested by Planck's theory of radiation.[27]

---

[22] Indeed, in his analysis Nicholson invented a hypothetical element which he called *Protofluorine*. Nicholson (1912), p. 679: "Protofluorine is one of the simplest forms of matter—that is to say, one of the simplest receptacles of energy which can exist." Nicholson (p. 677) assigned "a definite atomic constitution" to this contrived element.

[23] Bohr (1913), I, pp. 6, 22–23.

[24] Bohr (1913), I, p. 15. Cf. Nicholson (1912), p. 679.

[25] Bohr (1913), I, pp. 22–23. Cf. *BCW* 2 (1981), p. 109.

[26] Bohr (1913), I, p. 7; cf. p. 24.

Evidently, Bohr acknowledged the similarities between his theory and that of Nicholson.[28] Bohr did not call his own theory a model because, among other reasons, it included Planck's quantum of action; hence, for reason of consistency, he was unwilling to call Nicholson's theory a model. In brief, for Bohr the theories of Thomson and Rutherford were models, whereas his own theory and that of Nicholson were not.

Let us return to Thomson and Rutherford. In his essay on radioactive transformation of 1906 Rutherford writes under the heading "Representations of atomic constitution":

> The recent developments in physical science have given a great impetus to the study of the constitution of the atom, and attempts have been made to form a mechanical, or rather electrical, representation of an atom which shall imitate as closely as possible the behavior of the actual atom.[29]

So Rutherford thinks in terms of a mechanical or electrical representation of the atom's structure that can imitate the nature, that is, the constitution, of the actual atom: but a representation or an imitation is not the real thing. Rutherford mentions "constitution" in opposition to "representation", that is, the real thing is constitution—"the actual atom"—while representation refers to modeling. Rutherford does not do "modeling". He is very precise with his terminology:

> [J. J. Thomson's] model atom imitates in a remarkable way the behavior of the atom of the elements,.... Such attempts to imitate by an electrical model the structure of the atom are of necessity somewhat artificial, but they are of great value as indicating the general method of attack of the greatest problem that at present confronts the physicist.[30]

Rutherford clearly indicates that the model is a tool for pursuing research; it is a method, not the real thing.

In 1909, in his presidential capacity, Thomson reported to the *British Association for the Advancement of Science*:

> The interest inspired by equations ... in some minds is apt to be somewhat Platonic; and something more grossly mechanical—a model, for example, is felt by many to be more suggestive and manageable, and for them a more powerful instrument of research, than a purely analytical theory.[31]

---

[27] *BCW* 2 (1981), p. 109.

[28] *BCW* 2 (1981), p. 109. Bohr goes on to delineate differences between the two theories.

[29] Rutherford (1906), p. 263.

[30] Rutherford (1906), pp. 265–266.

[31] Thomson (1909), p. 268.

Thomson does not count himself as one of the "many" who consider model a powerful instrument. Rather, he presented his atomic theory in terms of structure.

This approach is already noticeable in 1904 when Thomson published his theory of the chemical atom, that is, the atom of the chemical elements, to which both Rutherford and Bohr later referred. In this paper Thomson plunges into many pages of computations on the stable arrangements of corpuscles within the atom which lead to the section, "Application of the preceding results to the theory of the structure of the atom."

> We suppose that the atom consists of a number of corpuscles moving about in a sphere of uniform positive electrification: the problems we have to solve are (1) what would be the structure of such an atom, *i.e.* how would the corpuscles arrange themselves in the sphere; and (2) what properties would this structure confer upon the atom.[32]

Thomson calculates that if the corpuscles were to arrange themselves in a series of concentric rings the structure would be stable, and he adds:

> I shall ... endeavour to show that the properties conferred on the atom by this ring structure are analogous in many respects to those possessed by the atoms of the chemical elements, and that in particular the properties of the atom will depend upon its atomic weight in a way very analogous to that expressed by the periodic law.[33]

This is Thomson's chemical atom, but nowhere in the article can one find the term "model". The key terms are "structures" and "systems" which "behave like...".[34] Thomson cannot deny the hypothetical character of his construction, but he consistently avoids using the term "model". Stability is achieved in analogous fashion; Thomson does not pretend to address the real chemical atom.

Let us now assess the situation. Bohr made a critical move, and a most productive one at that, from his earlier theoretical researches in Cambridge on the electron theory of metals to his study in Manchester of the atom. Already in July 1912, when he drafted the paper on the constitution of atoms and molecules, the so-called "Rutherford Memorandum", Bohr referred to the "atom-model proposed by Prof. Rutherford." He then continued to analyze the stability of "Thomson's atom-model".[35] This means that he considered both these theories atom-models. But when it came to his own theory, it was about constitution, not modeling.

---

[32] Thomson (1904a), p. 255.

[33] Thomson (1904a), pp. 255–256.

[34] Thomson (1904a), pp. 260, 262.

[35] *BCW* 2 (1981), p. 136. In fact, Bohr (p. 109) referred to Rutherford's atomic structure as a model in private correspondence with Rutherford in late January 1913.

Bohr primarily invokes the language of theory in which one makes assumptions that are intended to lead to consequences which are consistent with experimental evidence. He claims that these assumptions are supported by experimental data—which is the claim that the assumptions are well founded. His language does not suggest a model. For Bohr "constitution" (as in the title) indicates that he intends to describe the real thing (as it exists in nature):

> The inadequacy of the classical electrodynamics in accounting for the properties of atoms from an atom-model as Rutherford's, will appear very clearly if we consider a simple system consisting of a positively charged nucleus of very small dimensions and an electron describing closed orbits around it.[36]

Here Bohr distinguishes "model" from the actual state of nature:

> It is obvious that the behaviour of such a system will be very different from that of an atomic system occurring in nature. In the first place, the actual atoms in their permanent state seem to have absolutely fixed dimensions and frequencies.[37]

By "such a system" Bohr meant an atom which complies with the laws of classical electromagnetism, explicitly noting differences between the classical system and the way "actual atoms" behave.

In conclusion, it seems that Thomson, Rutherford, Nicholson, and Bohr all agreed that a model is a representation—distinct from reality—based on mechanical principles. Thomson and Rutherford believed they had in fact described reality, not a representation of it. Similarly, Bohr was convinced that he described reality; thus, for Bohr the term "model" is not appropriate for his own theory. According to Bohr, Thomson and Rutherford did not describe reality; hence, their theories are models (representations). Nicholson accepted the appropriateness of "model" to describe his own theory even though he was aware that Planck's quantum of action cannot be represented mechanically.

The expansion of the meaning of "model" to include Bohr's theory took place very quickly after the publication of Bohr's Trilogy. For Bohr modeling was inferior to a true description; he chose the terms "structure" and "constitution" which express, we submit, Bohr's clear intention to uncover the physical reality of the atom. But once Bohr's theory was acknowledged as representing the atom, it was implicitly accepted that modeling need not be entirely mechanical. Planck's quantum of action was an essential element of this model, that is, a discrete feature came to play a critical role in representing atomic phenomena.

---

[36] Bohr (1913), p. 3.

[37] Bohr (1913), p. 4.

**Acknowledgments.** We thank an anonymous referee for drawing our attention to a letter by Niels Bohr (n. 2). One of us [GH] is grateful to the Alexander von Humboldt Foundation for its support.

**Bibliography**

*BCW* 2 (1981). See Bohr (1981).

Bohr, Niels (1913). "On the constitution of atoms and molecules" [in three parts]. *Philosophical Magazine*, Series 6, 26, 1–25 [I], 476–502 [II], and 854–875 [III]; reprinted in Bohr (1981), pp. 161–233.

Bohr, Niels (1981). *Collected works.* Vol. 2: *Work on atomic physics (1912–1917),* U. Hoyer (ed.). Amsterdam and New York: North-Holland.

Heilbron, John L., and Thomas S. Kuhn (1969). "The genesis of the Bohr atom." *Historical Studies in the Physical Sciences* 1, 211–290.

Hon, Giora, and Bernard R. Goldstein (2012). "Maxwell's contrived analogy: An early version of the methodology of modeling." *Studies in History and Philosophy of Modern Physics* 43, 236–257.

Nicholson, John W. (1912). "The constitution of the solar corona, II." *Monthly Notices of the Royal Astronomical Society* 72, 677–692.

Rutherford, Ernest (1906). *Radioactive transformations.* New Haven: Yale University Press.

Rutherford, Ernest (1911). "The scattering of $\alpha$ and $\beta$ particles by matter and the structure of the atom." *Philosophical Magazine*, Series 6, 21, 669–688.

Sommerfeld, Arnold (1911). "Das Plancksche Wirkungsquantum und seine allgemeine Bedeutung für die Molekularphysik." *Physikalische Zeitschrift* 12, 1057–1069.

Sommerfeld, Arnold (1912). "Discussion du rapport de M. Planck." 115–132 in *La théorie du rayonnement et les quanta: Rapports et discussions de la réunion tenue à Bruxelles du 30 octobre au 3 novembre 1911 sous les auspices de M. E. Solvay.* Paris: Gauthier-Villars.

Thomson, Joseph J. (1904a). "On the structure of the atom: an investigation of the stability and periods of oscillation of a number of corpuscles arranged at equal intervals around the circumference of a circle; with application of the results to the theory of atomic structure." *Philosophical Magazine*, Ser. 6, 7, 237–265.

Thomson, Joseph J. (1904b). *Electricity and matter.* London: Archibald Constable.

Thomson, Joseph J. (1909). "Address of the President of the British Association for the Advancement of Science." *Science* 30 (27 Aug. 1909), 257–279.

Short biographies:

**Giora Hon** teaches History and Philosophy of Science at the University of Haifa, Israel. He has published widely on the theme of error in science, both from historical and philosophical points of view. His edited volume (together with M. Boumans, and A. Petersen), *Error and Uncertainty in Scientific Practice*, appeared recently (Pickering & Chatto, 2013). With Bernard R. Goldstein he coauthored the essay, "Maxwell's contrived analogy: An early version of the methodology of modeling," in *Studies in History and Philosophy of Modern Physics* (2012).

**Bernard R. Goldstein** is University Professor Emeritus at the University of Pittsburgh, USA. He has written extensively on various aspects of the history of science, including a series of papers jointly authored with Giora Hon, notably "How Einstein Made Asymmetry Disappear: Symmetry and Relativity in 1905," in *Archive for History of Exact Sciences* (2005).